\documentclass[twocolumn,showpacs,aps]{revtex4}
\usepackage[dvips]{graphicx}
\usepackage{bm}
\usepackage{mathrsfs}

\begin{document}
\date{\today}

\title{Curvature Dependent Diffusion Flow on Surface with Thickness}
\author{Naohisa Ogawa
\footnote{ogawanao@hit.ac.jp}}
\affiliation{Hokkaido Institute of Technology, Sapporo 006-8585 Japan}

\begin{abstract}
Particle diffusion in a two dimensional curved surface embedded in $R_3$ is considered.
In addition to the usual diffusion flow, we find a new flow with an explicit curvature dependence.
New diffusion equation is obtained in $\epsilon$ (thickness of surface) expansion.
As an example, the surface of elliptic cylinder is considered, 
and curvature dependent diffusion coefficient is calculated.
\end{abstract}
\maketitle

\section{Motivation}

The particle motion on a given curved surface is old but interesting problem in wide range of physics.
Especially the diffusion process of particles on such a manifold is still an open problem,
 and related to various kinds of phenomena.

For example the motion of protein on cell membrane has great importance in biophysics.
There are several research papers discussing on this problem.
Some of them are treating this problem by using usual diffusion equation with curved coordinate, 
and discuss the curvature (Gauss curvature) dependence of its solution \cite{diffusion_equation}.
Other of them use the Langevin equation on curved surface and calculating the curvature dependence of diffusion coefficient
 \cite{langevin_equation}.
 
 The quantum mechanics of particle motion on such a curved manifold is also considered by many authors.
This problem is usually explained by the Schroedinger equation  with Laplace-Beltrami operator.
However, when we treat the curved surface as embedded one in 3 dimensional Euclidean space,
 situation is changed and then we have a quantum potential term related to the curvature 
additional to the kinetic operator \cite{da Costa},\cite{ogawa_fujii},\cite{fujii}.

Another example is in larger scale physics in which our consideration is devoted.
Patterns of animal skins are well expressed by the reaction diffusion equation \cite{Turing}. 
But the patterns are different for each parts even in one individual. 
For example, Char fish, the side part has white spot pattern, but the back part has labyrinth pattern.
(For these two patterns, see for example \cite{Shoji_Iwasa}.)
One of the reasons might come from the curvature difference between side part and back part.
If the diffusion is influenced by the curvature, this difference of patterns might be explained.
 Furthermore, the cross section of fish has form of ellipsoid  and the surface can be approximated as the one of elliptic cylinder.
 In two dimensional space, we have only two kinds of curvature, 
one is Gauss curvature and other is mean curvature. 
Both are constructed from second fundamental tensor by taking determinant or trace.
Gauss curvature can also be constructed only by metric tensor and its derivatives, but this is not the case for the mean curvature.
The elliptic cylinder, in which we have much interest, has zero Gauss curvature and non-zero mean curvature.
Therefore to explain the pattern change of Char fish, solution of the diffusion equation should depend on mean curvature.
This is impossible if we start from usual diffusion equation because it depends only on metric but not on second fundamental tensor.
Therefore we need some new diffusion equation, which bring not only Gauss curvature but also mean curvature.
In this article, we discuss how to construct such curvature dependent diffusion equation.

\section{Coordinate and Metric}

The simple extension of diffusion equation in Euclidean space to Riemannian space can be done by
changing Laplacian with Cartesian coordinate to the one with curved coordinate, i.e. Laplace Beltrami operator.
This coordinate change is not enough for our purpose, however.
The way of construction of new diffusion equation in this paper is the followings.
We re-identify the two dimensional diffusion as the limiting process from three dimensional diffusion.
We place the curved surface $\Sigma$ in three dimensional Euclidean space $R_3$, 
and we put two similar copies of $\Sigma$, called $\tilde{\Sigma}$ and $\Sigma'$ at a small distance of $\epsilon/2$.
Our particles can only move between these two surfaces, and later we take a limit $\epsilon \to 0$.
We look for the form of diffusion equation in this limit.
The coordinates we use hereafter is the followings. (See fig.1)

$\vec{X}$ is the Cartesian coordinate in $R_3$.
$\vec{x}$ is the Cartesian coordinate which specifies only the points on $\Sigma$.
$q^i$ is the curved coordinate on $\Sigma$. (Small Latin indices $i,j,k,\cdots$ runs from 1 to 2.)
$q^0$ is the coordinate in $R_3$ normal to $\Sigma$.
Further by using the normal unit vector $\vec{n}(q^1,q^2)$ on $\Sigma$ at point $(q^1,q^2)$,
we can identify any points between two surfaces  $\Sigma'$ and $\tilde{\Sigma}$ 
by the following thin-layer approximation \cite{fujii}.

\begin{equation}
\vec{X}(q^0,q^1,q^2) = \vec{x}(q^1,q^2) + q^0 \vec{n}(q^1,q^2),
\end{equation}
where $ -\epsilon/2 \leq q^0 \leq \epsilon/2 $.
\begin{figure}
\centerline{\includegraphics[width=5cm]{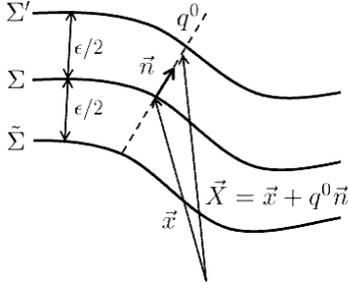}}
\caption{Embedding and Coordinate}
\end{figure}

From this relation we can obtain the curvilinear coordinate system between two surfaces ($\subset R_3$)
by the coordinate $q^{\mu}=(q^0,q^1,q^2)$, and metric $G_{\mu\nu}$. 
(Hereafter Greek indices $\mu, \nu,\cdots$ runs from 0 to 2.)

\begin{equation}
G_{\mu\nu}=\frac{\partial \vec{X}}{\partial q^{\mu}} \cdot \frac{\partial \vec{X}}{\partial q^{\nu}}.
\end{equation}

Each part of $G_{\mu\nu}$ is the following.

\begin{equation}
G_{ij}=g_{ij} + q^0 (\frac{\partial \vec{x}}{\partial q^{i}} \cdot \frac{\partial \vec{n}}{\partial q^{j}}
+ \frac{\partial \vec{x}}{\partial q^{j}} \cdot \frac{\partial \vec{n}}{\partial q^{i}}) + (q^0)^2 \frac{\partial \vec{n}}{\partial q^{i}} \cdot \frac{\partial \vec{n}}{\partial q^{j}},
\end{equation}
where
\begin{equation}
g_{ij} = \frac{\partial \vec{x}}{\partial q^{i}} \cdot \frac{\partial \vec{x}}{\partial q^{j}}
\end{equation}
is the metric on $\Sigma$. 
Hereafter indices $i,j,k \cdots$ are lowered or rised by $g_{ij}$ and its inverse $g^{ij}$.
We also obtain

\begin{equation}
G_{0i}=G_{i0}=0, ~~G_{00}=1.
\end{equation}

We can proceed the calculation by using the new variables.
We first define the tangential vector to $\Sigma$ by

\begin{equation}
\vec{B}_k = \frac{\partial \vec{x}}{\partial q^{k}}.
\end{equation}

Note that $\vec{n} \cdot \vec{B}_k =0$.
Then we obtain two relations.

Gauss equation: 
\begin{equation}
\frac{\partial \vec{B}_i}{\partial q^j} = - \kappa_{ij} \vec{n} + \Gamma^k_{ij} \vec{B}_k,
\end{equation}

Weingarten equation: 
\begin{equation}
\frac{\partial \vec{n}}{\partial q^j} = \kappa_j^m \vec{B}_m,
\end{equation}

where 
$$\Gamma^k_{ij} \equiv \frac{1}{2} g^{km}(\partial_i g_{mj} + \partial_j g_{im} - \partial_m g_{ij}).$$
$\kappa_{ij}$ is called Euler-Schauten tensor, or second fundamental tensor defined as

\begin{equation}
\kappa_{ij} = \frac{\partial \vec{n}}{\partial q^i}\cdot \vec{B}_j.
\end{equation}

The second fundamental tensor $\kappa_{ij}$ is the projection of $\partial \vec{n}$ into the surface.
Furthermore, the mean curvature is given by
\begin{equation}
\kappa = g^{ij} \kappa_{ij},
\end{equation}

and Ricci scalar curvature $R$ is obtained by

\begin{equation}
R/2 = \det(g^{ik} \kappa_{kj})= \det(\kappa^i_j) = \frac{1}{2}(\kappa^2 -\kappa_{ij} \kappa^{ij}).
\end{equation}

Then we have the formula for metric of curvilinear coordinate in a neighborhood of $\Sigma$.
\begin{equation}
G_{ij}=g_{ij} + 2 q^0 \kappa_{ij} + (q^0)^2 \kappa_{im} \kappa^m_j.
\end{equation}

Under the inversion $q^0 \to -q^0$,  we have $\kappa_{ij} \to - \kappa_{ij}$ as well as $\vec{n} \to -\vec{n}$ from $\vec{n}=\partial_0 \vec{X} / \mid \partial_0 \vec{X} \mid$.
Therefore $G_{ij}$ is invariant under $q^0 \to -q^0$.\\

Now we have the total metric tensor such as,
\begin{equation}
G_{\mu\nu} =
\left(
\begin{array}{cc}
1 & ~~~0~~~\\
0 &G_{ij}
\end{array}
\right).
\end{equation}

\section{Embedding of Diffusion field}

Let us denote 3 dimensional diffusion field as $\phi^{(3)}$, and Laplacian as $\Delta^{(3)}$.
Then we have the equation with normalization condition
\begin{eqnarray}
&& \frac{\partial \phi^{(3)}}{\partial t} = D \Delta^{(3)} \phi^{(3)},\label{eq:diff}\\
1 &=& \int \phi^{(3)}(q^0,q^1,q^2)  \sqrt{G}~ d^3 q,
\end{eqnarray}
where $D$ is the diffusion constant, and $G = \det(G_{\mu\nu}) = \det (G_{ij})$.
Our aim is to construct the effective two dimensional diffusion equation from 3D equation above.

\begin{eqnarray}
&& \frac{\partial \phi^{(2)}}{\partial t} = D \Delta^{(eff)} \phi^{(2)},\\
1 &=& \int \phi^{(2)}(q^1,q^2)  \sqrt{g}~ d^2 q,
\end{eqnarray}
where $\phi^{(2)}$ is the two dimensional diffusion field, $g = \det(g_{ij})$,
 and $\Delta^{(eff)}$ is unknown effective 2D diffusion operator which might not be equal to simple 2D Laplace Beltrami operator.

From two normalization conditions, we obtain
\begin{eqnarray}
1 &=& \int \phi^{(3)}(q^0,q^1,q^2)  \sqrt{G}~ d^3 q,\nonumber \\
&=& \int [\int_{-\epsilon/2}^{\epsilon/2} d q^0  (\phi^{(3)} \sqrt{G/g})] ~ \sqrt{g} ~ d^2 q, \nonumber\\
&=& \int \phi^{(2)}(q^1,q^2)  \sqrt{g}~ d^2 q.\nonumber
\end{eqnarray}
Therefore we obtain the relation,
\begin{equation}
\phi^{(2)}(q^1, q^2) = \int_{-\epsilon/2}^{\epsilon/2} \tilde{\phi}^{(3)} d q^0, \label{eq:relation}
\end{equation}
where 
\begin{equation}
\tilde{\phi}^{(3)} \equiv \phi^{(3)} \sqrt{G/g}.
\end{equation}

We multiply $\sqrt{G/g}$ to equation (\ref{eq:diff}) and integrate by $q^0$, then we obtain
\begin{equation}
\frac{\partial \phi^{(2)}}{\partial t} = D \int_{-\epsilon/2}^{\epsilon/2} \tilde{\Delta}^{(3)} \tilde{\phi}^{(3)} dq^0,\label{eq:laplace}
\end{equation}
where
\begin{equation}
 \tilde{\Delta}^{(3)} \equiv \sqrt{G/g} ~~ \Delta^{(3)}\sqrt{g/G}.
\end{equation}

Next we analyze new operator $\tilde{\Delta}^{(3)}$.
From the form of Laplace Beltrami operator
$$\Delta^{(3)} = G^{-1/2} \frac{\partial}{\partial q^\mu} G^{1/2} G^{\mu\nu} \frac{\partial}{\partial q^\nu},$$
we have
\begin{eqnarray}
\tilde{\Delta}^{(3)} &=& 
g^{-1/2} \frac{\partial}{\partial q^\mu} G^{1/2} G^{\mu\nu} \frac{\partial}{\partial q^\nu}(g/G)^{1/2}\nonumber\\
&=& \tilde{\Delta}^{(2)}  + \tilde{\Delta}^{(1)}, \label{eq:laplace2}
\end{eqnarray}
where
\begin{equation}
\tilde{\Delta}^{(2)} \equiv  g^{-1/2} \frac{\partial}{\partial q^i} G^{1/2} G^{ij} \frac{\partial}{\partial q^j}(g/G)^{1/2},
\end{equation}
and
\begin{equation}
\tilde{\Delta}^{(1)} \equiv \frac{\partial}{\partial q^0} G^{1/2} \frac{\partial}{\partial q^0}G^{-1/2}.
\end{equation}

Then our diffusion equation has form

\begin{equation}
\frac{\partial \phi^{(2)}}{\partial t} = D  \int_{-\epsilon/2}^{\epsilon/2} \tilde{\Delta}^{(2)} \tilde{\phi}^{(3)} dq^0. \label{eq:new}
\end{equation}

The contribution from $ \tilde{\Delta}^{(1)} $ vanishes because
\begin{eqnarray}
 \int_{-\epsilon/2}^{\epsilon/2} \tilde{\Delta}^{(1)} \tilde{\phi}^{(3)} d q^0 &=& g^{-1/2} \int_{-\epsilon/2}^{\epsilon/2} \frac{\partial}{\partial q^0} (G)^{1/2} \frac{\partial }{\partial q^0} \phi^{(3)} ~ dq^0  \nonumber\\
&=& g^{-1/2}  [~ (G)^{1/2} \frac{\partial \phi^{(3)}}{\partial q^0}] \mid _{-\epsilon/2}^{\epsilon/2} =0. 
\end{eqnarray}
The last equality is the requirement that diffusion flow does not pass through the surface: $\Sigma'$ and $\tilde{\Sigma}$.\\

Now we calculate r.h.s of (\ref{eq:new}) up to ${\cal O}(\epsilon^2)$.
Since we have
\begin{equation}
\tilde{\phi}^{(3)} = {\cal O}(\epsilon^{-1}),
\end{equation}
from (\ref{eq:relation}), we need to expand $\tilde{\Delta}^{(2)}$ up to ${\cal O}(\epsilon^2)$.
The following relations are useful

\begin{eqnarray}
G_{ij} &=& g_{ij} + 2 q^0 \kappa_{ij} + (q^0)^2 \kappa_{im} \kappa^m_j,\\
G^{ij} &=& g^{ij} - 2 q^0 \kappa^{ij} + 3 (q^0)^2 \kappa^i_{m} \kappa^{mj} + {\cal O}(\epsilon^3),\\
G_{~~} &=& g~ \{1 + 2 q^0 \kappa + (q^0)^2 (\kappa^2 + R) + {\cal O}(\epsilon^3)\},\\
G^{1/2}&=& g^{1/2} \{1 + q^0 \kappa + \frac{1}{2} (q^0)^2 R + {\cal O}(\epsilon^3)\},
\end{eqnarray}
where $R = \kappa^2 - \kappa_{ij} \kappa^{ij}$ is used.

Then the operator $\tilde{\Delta}^{(2)}$ can be expanded as follows

\begin{equation}
\tilde{\Delta}^{(2)}  = \Delta^{(2)}  + q^0 \hat{A} + (q^0)^2  \hat{B} +  {\cal O}(\epsilon^3),
\end{equation}
where,

\begin{equation}
\hat{A}  = - g^{-1/2} \frac{\partial}{\partial q^i} g^{1/2} (2 \kappa^{ij} \frac{\partial}{\partial q^j} + g^{ij} \frac{\partial \kappa }{\partial q^j}),
\end{equation}
\begin{eqnarray}
\hat{B}  &=& g^{-1/2} \frac{\partial}{\partial q^i} g^{1/2} (3 \kappa^{im} \kappa_m^j \frac{\partial}{\partial q^j} \nonumber\\
&+& \frac{1}{2} g^{ij} \frac{\partial (\kappa^2 -R) }{\partial q^j} + 2 \kappa^{ij} \frac{\partial \kappa }{\partial q^j} ).
\end{eqnarray}

Then our two dimensional effective diffusion equation up to ${\cal O}(\epsilon)$ is,
\begin{eqnarray}
\frac{\partial \phi^{(2)}}{\partial t} &=& D \Delta^{(2)} \phi^{(2)} \nonumber\\
&+& D \hat{A} \int_{-\epsilon/2}^{\epsilon/2} q^0 \tilde{\phi}^{(3)} dq^0 \nonumber\\
&+& D \hat{B} \int_{-\epsilon/2}^{\epsilon/2} (q^0)^2 \tilde{\phi}^{(3)} dq^0 + {\cal O}(\epsilon^3).
\end{eqnarray}

To proceed the $q^0$ integration, we suppose there is no diffusion flow in normal direction in layer , that is,
\begin{equation}
0 = \frac{\partial \phi^{(3)}}{\partial q^0} = g^{1/2} \frac{\partial G^{-1/2} \tilde{\phi}^{(3)}}{\partial q^0}.
\end{equation}
Solution is,
\begin{eqnarray}
\tilde{\phi}^{(3)} &=& \frac{1}{N}  (G/g)^{1/2} \phi^{(2)}(q^1,q^2),\\
N &\equiv & \int_{-\epsilon/2}^{\epsilon/2}  (G/g)^{1/2} dq^0.
\end{eqnarray}

Each integration can be explicitly performed, and we obtain
\begin{eqnarray}
N ~~~~~ &=& \epsilon +\frac{R}{24} \epsilon^3 + {\cal O}(\epsilon^5),\\
<q^0> ~ &=& \frac{\kappa_{} \epsilon^2}{12}  + {\cal O}(\epsilon^4),\\
<(q^0)^2> &=& \frac{\epsilon^2}{12}  + {\cal O}(\epsilon^4),
\end{eqnarray}
where we have used the definition
\begin{equation}
<f(q^0)> \equiv \frac{1}{N} \int_{-\epsilon/2}^{\epsilon/2} f(q^0) (G/g)^{1/2} dq^0.
\end{equation}

We obtain the final form of equation up to $ {\cal O}(\epsilon^2)$ as
\begin{flushleft}
\begin{eqnarray}
\frac{\partial \phi^{(2)}}{\partial t} &=& D \Delta^{(2)} \phi^{(2)} + \tilde{D} (\hat{A} \kappa + \hat{B}) \phi^{(2)}\nonumber \\
&=& D \Delta^{(2)} \phi^{(2)}  + \tilde{D} g^{-1/2} \frac{\partial}{\partial q^i} ~ g^{1/2} \nonumber\\
&\times& \{ (3 \kappa^{im} \kappa_m^j -2 \kappa \kappa^{ij}) \frac{\partial}{\partial q^j} 
- \frac{1}{2} g^{ij} \frac{\partial R}{\partial q^j} \} \phi^{(2)},~~~
\end{eqnarray}
\end{flushleft}
where $\tilde{D} = \frac{\epsilon^2}{12} D$.\\

We give two comments here.
First, ${\cal O}(\epsilon)$ term disappears. 
Since $\epsilon$ has the dimension of length, it always appears  with curvature $\kappa$.
Therefore the 1st order term, if it exists, it contains 1st order of curvature $\kappa$.
But this curvature depends on unphysical choice of normal unit vector $\vec{n}$, and so it does not appear.

Second, additional potential term disappears.
In quantum mechanics, the similar embedding techniques leads to the appearance 
of additional potential term written by curvature.
But in our classical case we have no such terms.
Because in diffusion equation, potential term breaks probability conservation law, i.e.
$$\partial \phi / \partial t =( D \Delta + V(x) )\phi, $$
$$ \to ~ \frac{d}{dt} \int d^3x ~\phi  = \int d^3x V(x) \phi \neq 0.$$

The normal diffusion flow can be written in general coordinate,
\begin{equation}
J_N^i = -D g^{ij} \frac{\partial \phi^{(2)}}{\partial q^j}.
\end{equation}
The anomalous diffusion flow is,
\begin{equation}
J_A^i = -\tilde{D} \{ (3 \kappa^{im} \kappa_m^j - 2\kappa \kappa^{ij})\frac{\partial \phi^{(2)}}{\partial q^j} 
- \frac{1}{2} g^{ij} \frac{\partial R }{\partial q^j}\phi^{(2)}\}. \label{eq:anom}
\end{equation}

The Diffusion equation is written as
\begin{eqnarray}
-\frac{\partial \phi^{(2)}}{\partial t} &=& \nabla_i (J_N^i + J_A^i),\nonumber\\
&=& g^{-1/2} \frac{\partial }{\partial q^j} ~g^{1/2} (J_N^i + J_A^i),
\end{eqnarray}
where $\nabla_i$ is the covariant derivative.
By using a suitable boundary condition, we can prove
$$ \frac{d}{dt} \int \phi^{(2)} g^{1/2} d^2 q =0.$$

\section{Properties of curvature dependent flow}
The anomalous flow equals to zero for the flat surface.
The last term in equation (\ref{eq:anom}) shows that curvature gradient generate the flow without particle density gradient.
From the signature of this term, this flow goes from the smaller Ricci scalar point to the larger Ricci scalar point.
Ricci scalar can take the negative, zero, and positive values. (Ricci scalar $R$ is related to Gauss curvature  by $R/2 = \det[\kappa^i_j]$.)
Let us work with the coordinate which satisfies $$g_{ij} = \delta_{ij},~~ \kappa^i_j = ~ \mbox{diag}[1/r_1,~1/r_2],$$ at point $P$, 
where $r_i$ is the curvature radius along the $q^i$ coordinate and it takes positive or negative value for convex or concave.
(The metric can be diagonalized by choosing the two coordinates as to satisfy orthogonality, and it can be normalized by using the re-parametrization.
The second fundamental tensor is diagonalized by rotation of coordinate.)

Then at point $P$, we have $R= \frac{2}{r_1r_2}$ and,

\begin{itemize}
 \item $R > 0$ ~~if  the surface is  convex or concave.
 \item $R = 0$ ~~if  the surface is essentially flat.
 \item $R < 0$ ~~if  the surface is hyperbolic.
\end{itemize}
Therefore the flow goes from hyperbolic or flat points to convex or concave points with positive 
larger Ricci scalar value.
\\

Next we consider the first term in (\ref{eq:anom}).
We have positive or negative value for
$$f^{ij} \equiv 3 \kappa^{im} \kappa_m^j - 2\kappa \kappa^{ij},$$
depending on the value of curvature.
In our coordinate, we can immediately write it in the simple form

\begin{equation}
f^{ij} = \delta^{ij} ( \frac{1}{(r_i)^2} - \frac{2}{r_1 r_2}).
\end{equation}

When the surface is hyperbolic ($R<0$), 
$$ f^{11} = \frac{1}{(r_1)^2} + \frac{2}{\mid r_1 r_2 \mid}>0,~~ 
f^{22} = \frac{1}{(r_2)^2} + \frac{2}{\mid r_1 r_2 \mid}>0,$$
usual diffusion occurs (See fig. 2).

\begin{figure}
\centerline{\includegraphics[width=4cm]{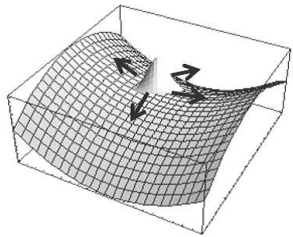}}
\caption{Wave packet on hyperbolic surface diffuses in two directions.}
\end{figure}

When the surface is convex or concave ($R>0$),
$$ f^{11} = \frac{1}{(r_1)^2} - \frac{2}{\mid r_1 r_2 \mid} \\
= \frac{\mid r_2\mid -2 \mid r_1\mid}{\mid r_1\mid^2 \mid r_2\mid},$$
$$f^{22} = \frac{1}{(r_2)^2} - \frac{2}{\mid r_1 r_2 \mid}\\
= \frac{\mid r_1\mid -2 \mid r_2\mid}{\mid r_2\mid^2 \mid r_1\mid}.$$

In this case, we have three possibilities.
One possibility is that both are negative, if 
$$1 /2 < \mid \frac{r_2}{r_1}\mid < 2.$$
Then we have no diffusion but concentration occurs (See fig. 3).

\begin{figure}
\centerline{\includegraphics[width=4cm]{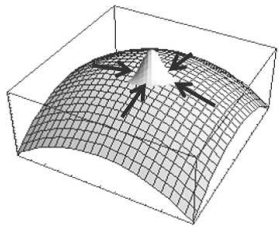}}
\caption{Wave packet on convex surface concentrates.}
\end{figure}

The second possibility is that one of two is positive and the other is negative, 
if  $$\mid \frac{r_2}{r_1}\mid < 1/2, ~~ \mbox{or} ~\mid \frac{r_2}{r_1}\mid > 2.$$
Then we have diffusion in one direction, but concentration in another direction (See fig. 4).

The third possibility for $R>0$ is, 

$$ \mid \frac{r_2}{r_1}\mid = 1/2, ~~ \mbox{or} ~ \mid \frac{r_2}{r_1}\mid =2. $$
This is critical point, where the flow stops for larger curvature direction
 and flow concentrates for smaller curvature direction.
\\

\begin{figure}
\centerline{\includegraphics[width=4cm]{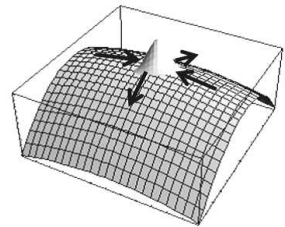}}
\caption{Wave packet on convex surface with one direction curvature is over two times higher than another. 
Packet diffuses in higher curvature direction and concentrates in smaller curvature direction.}
\end{figure}

When the Ricci scalar is zero ($R =0$), for example $r_2 = \infty$, $f^{22}=0$ and $f^{11}>0$ follows.
The diffusion occurs only in $q^1$ direction but not in another direction. (See fig. 5).

\begin{figure}
\centerline{\includegraphics[width=4cm]{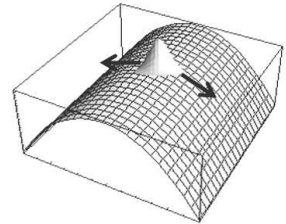}}
\caption{Wave packet on elliptic cylinder.
Packet diffuses only in curved direction but not in another direction.}
\end{figure}

In this way, this anomalous diffusion flow has much varieties depending on the curvature.
\\

\section{One Example: Elliptic Cylinder}
Let us consider one simple example where diffusion coefficient depends on curvature. 
The surface of elliptic cylinder is the case of $R=0$ just as figure 5, but the surface has non zero mean curvature.

Ellipsoid is given by the equation
\begin{equation}
 (\frac{x}{a})^2 + (\frac{y}{b})^2  = 1 .
\end{equation}

\begin{figure}
\centerline{\includegraphics[width=3cm]{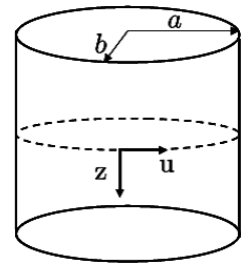}}
\caption{Elliptic cylinder}
\end{figure}

Then any points on cylinder  are specified by curved coordinate  $\theta$ and $z$;
\begin{equation}
x = a \cos \theta, ~~ y = b \sin \theta.
\end{equation}

Another choice of coordinate instead of  $\theta$ is,
\begin{equation}
du = \sqrt{dx^2+ dy^2} =  f(\theta) ~d\theta.
\end{equation}
where $f(\theta)$ is defined as
\begin{equation}
f(\theta) \equiv \sqrt{ a^2 \sin^2\theta + b^2 \cos^2 \theta} .
\end{equation}

The length of $u$ is given by
\begin{eqnarray}
u(\phi) &=& \int_0^{\phi} f(\theta) d\theta \nonumber\\
&=& b \int_0^{\phi} \sqrt{1-k^2 \sin^2 \theta} d\theta \equiv b E(k,\phi),\label{eq:u}
\end{eqnarray}
with $k= \sqrt{1-(a/b)^2},~ a \leq b.$
$E(k,\phi)$ is the Elliptic integral of the second kind. ~(See appendix)\\

The total length of $u$ is given by 
$$U \equiv 4b E(k, \pi/2).$$
We use the normalized value for $u$ hereafter such that,
\begin{equation}
\tilde{u} = u/U, ~~ 0 \leq \tilde{u} \leq 1.
\end{equation}

The normal unit vector $\vec{n}$ is given as
\begin{eqnarray}
\vec{n} &=&  (\frac{x}{a^2 \sqrt{(x^2/a^4) + (y^2/b^4)}}, \frac{y}{b^2 \sqrt{(x^2/a^4) + (y^2/b^4)}})\nonumber\\
&=& \frac{1}{f(\theta)} (b \cos \theta, a \sin \theta).
\end{eqnarray}

\begin{equation}
\frac{\partial \vec{n}}{\partial \theta} = \frac{1}{f} (-b \sin \theta, a \cos \theta) - \frac{\partial_\theta f}{f} \vec{n}.
\end{equation}
\begin{equation}
\vec{B}_\theta = \frac{\partial \vec{x}}{\partial \theta} = (-a \sin \theta, b \cos \theta).
\end{equation}
Then we obtain the second fundamental tensor.
\begin{equation}
\kappa_{\theta\theta} =\vec{B}_\theta \cdot \frac{\partial \vec{n}}{\partial \theta} = \frac{ab}{f} .
\end{equation}
Then we collect all the necessary quantities as follows
\begin{eqnarray}
&& g_{\theta\theta}=f^2, ~~ g_{zz}=1, ~~ g_{\theta z}=0, \nonumber\\
&& \kappa_{\theta\theta} =  \frac{ab}{f}, ~~\kappa_{zz} = \kappa_{\theta z}=0, ~~ \kappa = \frac{ab}{f^3}. \label{eq:mean}
\end{eqnarray}

Then we obtain the total diffusion equation expressed by the parameters $\theta$ and  $z$.
\begin{equation}
\frac{\partial \phi^{(2)}}{\partial t} =  (\frac{1}{f}\frac{\partial}{\partial \theta})  D_{\theta} (\frac{1}{f}\frac{\partial}{\partial \theta}) \phi^{(2)}  + D \frac{\partial^2}{\partial z^2} \phi^{(2)},
\end{equation}
where the effective diffusion coefficient depends on mean curvature.
$$D_{\theta} = D (1 + \frac{\epsilon^2 \kappa^2}{12}) = D (1 + \varepsilon^2 (b\kappa)^2) ,$$
where $\varepsilon \equiv  \epsilon /(2\sqrt{3}b)$.

Under this equation, we obtain the following particle number conservation law.
$$ \frac{d}{dt} \int dz \int_0^{2\pi} d\theta ~f(\theta)~ \phi^{(2)}(\theta, z)  ~= 0$$
with suitable Neumann boundary condition.

By using the variable $\tilde{u}$ instead of $\theta$ we have simple dimensionless equation,
\begin{equation}
\frac{\partial \phi^{(2)}}{\partial \tau} =  \frac{\partial}{\partial \tilde{u}} (1 + V)  \frac{\partial}{\partial \tilde{u}} \phi^{(2)}  + \frac{\partial^2}{\partial \eta^2} \phi^{(2)}
\end{equation}
where $ \tau = t D/U^2, ~\eta = z/U, ~ V = \varepsilon^2 (b\kappa)^2, ~ U= 4b E(k,\pi/2)$.\\

By using the approximation of elliptic function given in appendix, curvature dependent potential $V$ can be written as function of $\tilde{u}$.
The simulation can be done as usual diffusion equation.
For $0 \leq \tilde{u} \leq 1$ and $0 \leq \eta \leq 4$ using periodic boundary condition, 
since the length of $\eta$ is larger than one of $\tilde{u}$, the diffusion in $u$ direction occurs fastly 
and then the diffusion in $\eta$ direction follows slowly just like $\phi \sim a + \sum_k b(k) e^{- k^2 \tau} \cos(k \eta)$.
The $u$-directional diffusion can not occur uniformly, because at $\tilde{u}=0.25,$ and $0.75$ the diffusive coefficient is higher than other points. Therefore the slope of diffusion field is small especially at these two points during the diffusion process. (fig. 7)

\begin{figure}
\centerline{\includegraphics[width=9cm]{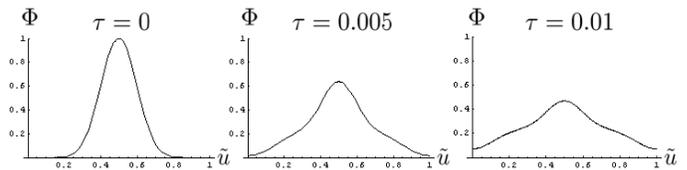}}
\caption{Snap shot of diffusion process starting from the wave packet 
 $\phi = \sin^{10} (\pi \tilde{u})$ as the initial condition. 
At two points (0.25 and 0.75), diffusion occurs quickly and its slope is smaller than other.}
\end{figure}

\begin{figure}
\centerline{\includegraphics[width=4cm]{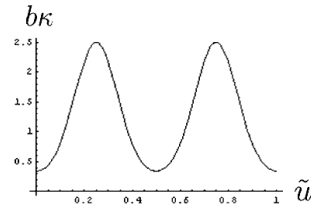}}
\caption{Mean curvature as a function of $\tilde{u}$ when $b/a=2$}
\end{figure}

\section{Conclusion}
 We have discussed on the diffusion equation on curved surface embedded in $R_3$.
We obtained the new diffusion equation up to  ${\cal O}(\epsilon^2)$ 
which includes anomalous diffusive flow additional to the usual one.
This anomalous flow depends on the second fundamental tensor, 
and it has not only diffusion but also concentration properties depending on the curvature of its manifold.

At the point with negative Ricci scalar $R<0$, surface is hyperbolic, and diffusion in both direction occurs. (fig.2) 

When Ricci scalar is positive $R>0$, we have three possibilities. 
$r_i$ appearing below is curvature radius in each direction ($i=1, 2$).

\begin{itemize}
\item
Concentration in both direction (fig.3), when 
$$1 /2 < \mid \frac{r_2}{r_1}\mid < 2.$$ 
\item
Concentration in smaller curvature direction, and diffusion in higher curvature direction (fig.4), when
$$\mid \frac{r_2}{r_1}\mid < 1/2, ~~ \mbox{or} ~\mid \frac{r_2}{r_1}\mid > 2.$$ 

\item
Concentration in smaller curvature direction, and no flow in higher curvature direction, when
$$\mid \frac{r_2}{r_1}\mid = 1/2, ~~ \mbox{or} ~\mid \frac{r_2}{r_1}\mid = 2.$$ 
\end{itemize}

When Ricci scalar is zero $R=0$, surface is essentially flat, but we have finite curvature radius in one direction.
Then we have diffusion only in this direction (fig.5).

 In the case of surface of elliptic cylinder, we gave a concrete form of equation
and we showed the curvature dependent diffusion coefficient.

$$ D_{u} = D (1 + \frac{\epsilon^2 \kappa^2}{12}), ~~~ D_{z} = D, $$
where $\kappa$ is the mean curvature.
In this case curvature dependence is simply included into diffusion coefficient. 
However this is not true in general case,
where situation is much more complicated, and this can be seen from the form of anomalous flow.

The application to the pattern formation by reaction diffusion using this obtained equation is not yet finished.
This will be done in further publication.

\section{Appendix}
We approximate the elliptic integral of the second kind.

\begin{equation}
E(k,\phi) \equiv \int_0^\phi \sqrt{1-k^2 \sin^2 \theta} ~ d\theta
\end{equation}
with
$$k = \sqrt{1-(a/b)^2}.$$

Under the expansion in powers of $k^2$, we obtain the power series of Elliptic integral of the second kind.

\begin{equation}
E(k,\phi) = \phi - \sum_{n=1}^{\infty} \frac{k^{2n} (2n-3)!!}{n! ~ 2^n} \int_0^\phi \sin^{2n}\theta d\theta.
\end{equation}

Since the integration part can be expanded by $\phi$ and $\sin 2n \phi$, we have

\begin{equation}
E(k,\phi) = a_0 \phi + \sum_{n=1}^{\infty} a_n \sin 2n \phi.
\end{equation}
with the relation
\begin{eqnarray}
a_0 &=& \frac{2}{\pi} E(k,\pi/2),\\
a_n &=& (-1)^n \frac{2E(k,\pi/2)}{n \pi} \nonumber\\
&& + \frac{4}{\pi} \int_0^{\pi/2} \sin 2n\phi ~E(k,\phi) d\phi. ~~(n \geq 1)
\end{eqnarray}

For the real Char fishes, $b/a$ takes values $1.5 \sim 2.5$. Then the value of $k$ takes $0.75 \sim 0.92$.
When $b/a=2$, each values of $a_n$ is given numerically
$$a_0=0.771,~~a_1=0.123,~~a_2=-0.00506,~~a_3=0.000558.$$

Now we have for $u$ given in (\ref{eq:u}),

\begin{equation}
u/b = E(k,\phi) = a_0 \phi + a_1 \sin 2\phi + a_2 \sin 4\phi + \cdots.
\end{equation}

And we rewrite it by using normalized $u$,

\begin{equation}
\phi = 2\pi \tilde{u} - \frac{a_1}{a_0} \sin 2\phi - \frac{a_2}{a_0} \sin 4\phi - \cdots,
\end{equation}
where $\tilde{u} = u/(4b E(k, \pi/2)).$

The iteration method up to order $(a_1/a_0)^1$ gives

\begin{equation}
\phi = 2\pi \tilde{u} - \frac{a_1}{a_0} \sin (4 \pi \tilde{u}).
\end{equation}

Then we take the derivative by $u$ in both hand sides.

\begin{equation}
\frac{1}{f} = \frac{1}{b a_0} (1 - \frac{2 a_1}{a_0} \cos (4 \pi \tilde{u})),
\end{equation}
where the following relation is used.
$$\frac{du}{d\phi} = f(\phi) \equiv \sqrt{a^2 \sin^2\phi + b^2 \cos^2\phi}.$$

Then the mean curvature given by (\ref{eq:mean}) is obtained as function of $u$.

\begin{equation}
b \kappa = \frac{a b^2}{f^3} = \frac{1}{\beta (a_0)^3} (1 - \frac{2 a_1}{a_0} \cos (4 \pi \tilde{u}))^3,
\end{equation}
where $\beta = b/a$.
This function is shown in figure 8.

\end{document}